\documentclass[%
 reprint,
 amsmath,amssymb,
 aps,
]{revtex4-1}

\usepackage{url}
\usepackage{dcolumn}
\usepackage{bm}

 \usepackage{graphicx}

\usepackage{amsbsy}

\begin{document}




\title{Tracking fast neutrons}


\author{Zhehui Wang}
\email{zwang@lanl.gov}
\author{Christopher L. Morris}%
\email{cmorris@lanl.gov}
\affiliation{Los Alamos National Laboratory, Los Alamos, NM 87545}

\date{\today}

\begin{abstract}
Based on elastic collisions, the linear momentum of a fast neutron can be measured from as few as two consecutive recoil ion tracks plus the vertex position of the third collision, or `two and half' ion tracks. If the time delay between the first two consecutive ion tracks is also measured, the number of ion tracks can be reduced to one and a half. The angular and magnitude resolutions are limited by ion range straggling to about ten percent.  Multi-wire proportional chambers and light-field imaging are discussed for fast neutron tracking. Single-charge or single-photon detection sensitivity is required in either approach. Light-field imaging is free of charge-diffusion-induced image blur, but the limited number of photons available can be a challenge.  $^1$H,$^2$H and $^3$He could be used for the initial development of fast neutron trackers based on light-field imaging. 

\end{abstract}


\maketitle




\section{Introduction}
\label{sec:1}
Recently, several experiments were reported that  simultaneously measured the directions and energies of fast neutrons from recoil ion tracks~\cite{Hunter2008,Roccaro:2009,Jovanovic:2009,Lopez:2011}. In the experiments, $^1$H or $^4$He nuclei were used as the ion targets, the ionization tracks of MeV recoil protons or $\alpha$-particles were recorded using either electronic or optical readouts. In addition to fast neutron counting, the directions and energies of fast neutrons were inferred from the ion tracks. The collisions were predominantly elastic.  A neutron-tracking scheme based on $^3$He absorption was also described~\cite{Hunter2008}.

It is well known that a single elastic encounter between a neutron and a nucleus is insufficient to transfer all of the linear or translational momentum of the neutron to the nucleus, except for the rare occasion of a head-on collision between a neutron and a proton~\cite{BC36,Fermi49,Marshak47,GE:52,WW:58,Knoll2000}. Here we show analytically that, based on elastic collisions, the linear momentum of a fast neutron can be measured from as few as three consecutive elastic encounters between the neutron and a low-Z target consisting of $^1$H, $^2$H or $^4$He atoms. The effects of an electromagnetic field or gravity on fast neutron motion can be ignored here. A fast neutron trajectory is a line segment in between two consecutive collisions, which can be determined by localizing the vertices of the collisions. The use of a higher-Z target, such as $^{12}$C, is hindered by low detector sensitivity to ions with energies below 100 keV. The energy transfer from a neutron decreases when the atomic number $A$ of the target increases~\cite{Fermi49,Marshak47,GE:52,WW:58}.

 Besides a head-on elastic collision with a proton, a fast neutron can also transfer its full linear momentum to a target nucleus in a single collision when it is absorbed. $^3$He, $^6$Li and $^{10}$B are of interest because of their relatively small masses~\cite{Knoll2000}. In each case, the corresponding compound nuclei $^4$He*, $^7$Li*, and $^{11}$B*  are not stable and disintegrate into two ions that can be tracked, and the fast neutron momentum can be deduced. In the case of $^{11}$B*, 94\% of the decay branch also emits a Doppler-shift $\gamma$-ray, which may be used to obtain neutron momentum. The neutron absorption cross sections are less than 1 barn for MeV neutrons, high detection efficiency would require a large amount of target mass. Indeed, neutron absorption by $^3$He, $^6$Li or $^{10}$B is widely used for thermal neutron detection since the absorption cross sections scale inversely with the speed of a neutron~\cite{Knoll2000,Fraga:2002,Runkle:2010}.  Although the present discussion of fast neutron tracking primarily focuses  on elastic collisions, the measurement techniques below should also be applicable to absorption-based neutron tracking. 
 
Below, we first describe fast neutron tracking principle in Sec.~\ref{princ:1}. Next, we show that the measurement errors are limited by ion range straggling. In Sec.~\ref{sec:TM}, we discuss neutron tracking using multi-wire proportional chambers and light-field imaging. In the following section, comparisons between elastic collision based methods and absorption based methods are made. A brief summary is given towards the end.

\section{Neutron-tracking principle}
\label{princ:1}
A fast neutron (MeV in kinetic energy) moves in a straight line and makes a sudden turn occasionally when it gets in a close proximity to $\sim$ 10$^{-15}$ m of another nucleus. For a low-Z nucleus like  $^1$H, $^2$H, $^3$He, $^6$Li, or $^{10}$B, elastic collision and absorption are most likely during the close encounter. The details of nuclear interaction are not needed for an elastic collision and the trajectories of the scattered neutron and the recoil nucleus (it is used with `recoil ion' interchangeably) can be determined from the conservation of momentum and energy within the framework of classical mechanics~\cite{Fermi49,Marshak47,GE:52,WW:58}. The momentum of the scattered neutron, as well as the momentum of the recoil nucleus, is random. The angular distributions of the scattered momenta do depend on the details of nuclear interaction, such as the type of nucleus, nuclear spin and neutron energy.

\begin{figure}[thbp] 
   \centering
   \includegraphics[width=2in]{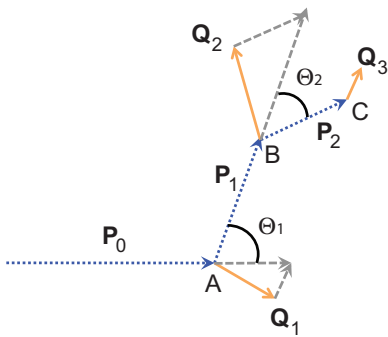} 
   \caption{Fast neutron linear momentum measurement based on three recoil ion tracks. A, B and C are the three vertices of three consecutive elastic collisions. The ion momenta are {\bf Q}$_1$, {\bf Q}$_2$, and {\bf Q}$_3$. The initial neutron momentum is {\bf P}$_0$. The neutron momentum after the $i$th collision is labeled as {\bf P}$_i$. {\bf Q}$_1$, {\bf Q}$_2$, and {\bf Q}$_3$ are in general not on the same plane. But the triplets $\{${\bf P}$_0$, {\bf P}$_1$, {\bf Q}$_1\}$ and $\{${\bf P}$_1$, {\bf P}$_2$, {\bf Q}$_2\}$ are each coplanar because of momentum conservation.}
   \label{fig:1}
\end{figure}

 In the first scenario, which solely relies on the recorded positions of recoil ion tracks, three ion tracks are sufficient to measure the linear momentum of the fast neutron. We shall assume that the track lengths can be used to derive recoil ion energies. The assumption is valid to within the straggling ranges of ion stopping. The straggling ranges are within a few percent of the mean stopping ranges of MeV ions. In Fig.~\ref{fig:1}, the measured quantities are recoil ion momenta {\bf Q}$_1$, {\bf Q}$_2$, and {\bf Q}$_3$. The initial neutron momentum is {\bf P}$_0$. The subsequent neutron momenta are labelled as {\bf P}$_1$ and {\bf P}$_2$. Momentum conservation gives
\begin{equation}
{\bf P}_0 = {\bf Q}_1 + {\bf P}_1, 
\label{sum:1}
\end{equation}
with
\begin{equation}
{\bf P}_1 = {\bf Q}_2 + {\bf P}_2. 
\label{sum:2}
\end{equation}

The directions of {\bf P}$_1$ and {\bf P}$_2$, symbolized here by the unit vectors {\bf e}$_1$ and {\bf e}$_2$, can be obtained from the locations of the vertices A, B and C. The momenta satisfy {\bf P}$_1 =$ p$_1$ {\bf e}$_1$ and {\bf P}$_2 =$ p$_2$ {\bf e}$_2$ respectively with $p_1$ and $p_2$ being their magnitudes. The dot product of {\bf e}$_1$ with Eq.~(\ref{sum:2}) is
\begin{equation}
p_1 = {\bf Q}_2 \cdot {\bf e}_1 + p_2 {\bf e_1}\cdot {\bf e}_2.
\label{proj:1}
\end{equation}
The dot product of {\bf e}$_2$ with Eq.~(\ref{sum:2}) is
\begin{equation}
p_1 {\bf e_1}\cdot {\bf e}_2 = {\bf Q}_2 \cdot {\bf e}_2 + p_2. 
\label{proj:2}
\end{equation}
Solving Eqs.~(\ref{proj:1}) and (\ref{proj:2}) for $p_1$ and $p_2$, 
\begin{equation}
p_1 = \frac{{\bf Q}_2 \cdot {\bf e}_1 -({\bf Q}_2 \cdot {\bf e}_2)({\bf e_1}\cdot {\bf e}_2) }{1- ({\bf e_1}\cdot {\bf e}_2 )^2},
\label{p1:res1}
\end{equation}
and
\begin{equation}
p_2 = - \frac{{\bf Q}_2 \cdot {\bf e}_2 -({\bf Q}_2 \cdot {\bf e}_1)({\bf e_1}\cdot {\bf e}_2) }{1- ({\bf e_1}\cdot {\bf e}_2 )^2}.
\label{p2:res1}
\end{equation}
It is clear that the full recording of the third ion track {\bf Q}$_3$ is not necessary. The collision vertex C, together with the location of B, is suffice to obtain {\bf e}$_2$. In short, the vectors {\bf Q}$_1$, {\bf Q}$_2$, {\bf e}$_1$ and {\bf e}$_2$, or `2 and half' ion tracks, are sufficient to determine {\bf P}$_0$. If one approximates the ion tracks by straight lines, the set $\{${\bf Q}$_1$, {\bf Q}$_2$, {\bf e}$_1$, {\bf e}$_2\}$ is derivable from a five-point measurement, the three vertices A, B and C, together with the end points of the first and second recoil ion track.

In the second scenario, the five-point measurement can be reduced to a three-point measurement, if the neutron time-of-flight (TOF) can be measured in between ion tracks. Specifically, if the time delay between the  first (point A in Fig.~\ref{fig:1}) and the second (point B) collision, symbolized by $\tau_1$, is measured, then $\tau_1$ uniquely determines the magnitude $p_1$ and the third collision is not necessary. Only the vertices of the first and second collisions are needed to deduce {\bf e}$_1$. The set  $\{${\bf Q}$_1$, {\bf e}$_1$, $\tau_1$$\}$ determines {\bf P}$_0$ through Eq.~(\ref{sum:1}).

One may ask whether consecutive Compton scatterings of gamma ($\gamma$) rays can be used for tracking $\gamma$ rays. The fact that $\gamma$ rays always move at the speed of light necessitates the five-point measurement.  The stopping powers for Compton electrons are smaller than those of ions in the MeV energy range. Electron tracks can not be approximated by straight lines. These differences make $\gamma$-ray tracking unique in its own way and $\gamma$-ray tracking will not be discussed further. 

\section{Measurement Errors} 
\label{sec:ME}
The error in a vector measurement can be broken down to the error in magnitude and the error in direction, $\delta${\bf K}$_0$ = $\delta $k$_0${\bf e}$_0$ + k$_0 \delta${\bf e}$_0$. Here {\bf K}$_0 = k_0 {\bf e}_0$ is the vector. $\delta $k$_0${\bf e}$_0$ describes the error in magnitude and k$_0 \delta${\bf e}$_0$ the error in direction, since $\delta${\bf e}$_0$$\cdot${\bf e}$_0$ = 0. For example, for a length vector based on the measurement of two end points, {\bf r}$_0$ = {\bf r}$_2$ - {\bf r}$_1$, the relative error in length ($\epsilon_r$) is given by
\begin{equation}
\epsilon_r \equiv \frac{\sqrt{\delta r_0^2}}{r_0} = \frac{\sqrt{2} \delta l}{r_0},
\label{raderr:1}
\end{equation}
where we assume that the errors in the three coordinates are the same $\sim \delta l$ for both ends {\bf r}$_1$($x_1$, $y_1$, $z_1$) and {\bf r}$_2$($x_2$, $y_2$, $z_2$). That is, $\delta x_1 = \delta y_1 = \cdots = \delta z_2 = \delta l$.
The error in direction ($\epsilon_\phi$) is given by
\begin{equation}
\epsilon_\phi = \frac{|r_0 \delta {\bf e}_0|}{r_0}=\sqrt{\delta {\bf e}_0 \cdot \delta {\bf e}_0}.
\end{equation}
It is straightforward to show that
\begin{equation}
\epsilon_\phi = \frac{2 \delta l}{r_0}.
\label{phierr:1}
\end{equation}

More generally, if the errors in different coordinates vary, one may define
\begin{equation}
\delta l^2 \equiv \max [\frac{\delta x_2^2 + \delta x_1^2}{2}, \frac{\delta y_2^2 + \delta y_1^2}{2}, \frac{\delta z_2^2 + \delta z_1^2}{2} ],
\end{equation}
then Eqs.~(\ref{raderr:1}) and (\ref{phierr:1}) are valid as the upper bounds on the errors using the generalized definition of $\delta l$.

The experimental implications are that a.) the ratio of an ion track length to the spatial resolution of a track recording method determines the errors in magnitude and direction of an ion momentum {\bf Q}$_i$ and b.) $\delta l \ll r_0$ is desirable for small errors. 

\begin{figure}[thbp] 
   \centering
   \includegraphics[width=3.0in]{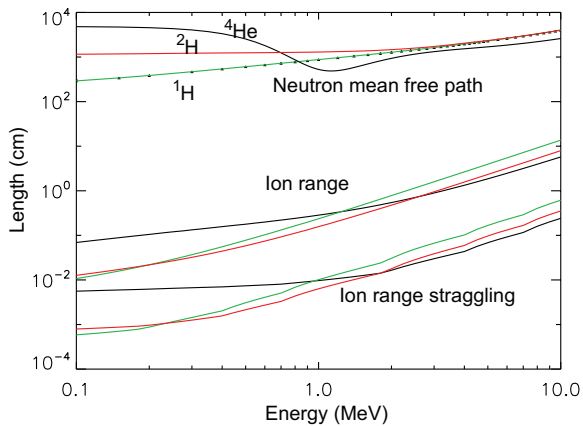} 
   \caption{A natural length hierarchy exists in the fast neutron tracking problem, $\lambda \gg L \gg \delta L_{sl} \sim \delta L_{st}$, with $\lambda$ being the mean-free-path of neutron elastic collision, $L$ the recoil ion range, $\delta L_{sl}$ being the longitudinal and  $\delta L_{st}$ the lateral (`transverse') ion straggling ranges respectively. The calculations are for ${^1}$H, ${^2}$H, ${^4}$He with a molecular density of 2.69$\times$ 10$^{20}$ per cm$^{3}$ (or 10 bar pressure at STP). The ratios of the lengths are approximately invariant to density changes. The neutron mean free paths are based on cross sections given by~\cite{t2:lanl}. The ion ranges and range straggling are outputs from SRIM calculations~\cite{SRIM}.}
   \label{fig:scaling}
\end{figure}

\begin{figure}[thbp] 
   \centering
   \includegraphics[width=2.0in]{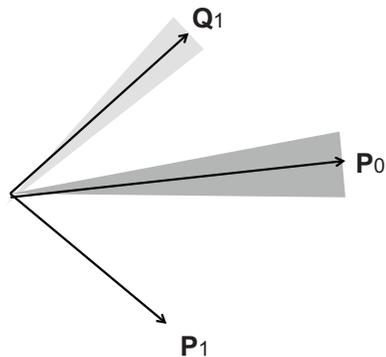} 
   \caption{Because fast neutron mean free path is much greater than recoil ion range, a good approximation is that the errors in the directions of neutron momenta after the first collision can be neglected. Therefore, the error in the incoming neutron momentum ${\bf P}_0$ comes from both the directional and magnitude uncertainties in ${\bf Q}_1$ and only the magnitude error in {\bf P}$_1$.}
   \label{fig:err}
\end{figure}

In the fast neutron tracking problem, fortunately, there is a natural length scaling, $\lambda \gg L \gg \delta L_{sl} \sim \delta L_{st}$, as shown in Fig.~\ref{fig:scaling}. Here $\lambda$ is the mean-free-path of neutron elastic collision. $L$ is the recoil ion range. $\delta L_{sl}$ is the ion longitudinal straggling range  and  $\delta L_{st}$ the lateral  or `transverse' ion straggling range. This length scaling implies that  errors in neutron directions, {\bf P}$_1$ and ${\bf P}_2$ in Fig.~\ref{fig:1}, which are proportional to $\delta l/ \lambda $ (much less than 1\%),  are much smaller than the errors in ion directions, {\bf Q}$_1$ and {\bf Q}$_2$, which are proportional to $\delta l/ L$ (around 10\%). Here $\delta l$ stands for the detector spatial resolution. In the analysis below, we will neglect the direction errors in neutron momenta after the first collision, Fig.~\ref{fig:err}. The error in the neutron momentum ${\bf P}_0$ comes from the directional and magnitude uncertainties in ${\bf Q}_1$ and the magnitude uncertainty in {\bf P}$_1$.

We begin by examining the error in the first recoil ion momentum, $\delta${\bf Q}$_1$, which can be determined from two positions, the vertex A and the end point of the recoil ion track.
\begin{equation}
\delta{\bf Q}_1 = \delta q_1 {\bf f}_1 + q_1 \delta {\bf f}_1,
\end{equation}
here we use {\bf Q}$_1  \equiv q_1${\bf f}$_1$ and {\bf f}$_1$ is the unit vector for the ion momentum {\bf Q}$_1 $.
The momentum of a recoil ion is related to its energy through 
\begin{equation}
\delta q_1 = \frac{\delta E_1}{\beta_1 c},
\end{equation}
with $E_1 = (\gamma_1 -1)m_1c^2 $. Here we keep the relativistic formulation for the ion. $\beta_1$ is the ratio of the ion speed to the speed of light $c$, and so forth. Since $E_1=E_1(L_1)$, the energy is a function of the ion range, $\delta E_1 = \delta L_1 (dE_1/dL_1) $. Therefore,
\begin{equation}
\delta q_1 = \frac{\delta L_1}{\beta_1 c} \frac{dE_1}{dL_1},
\end{equation}

Two factors contribute to the ion range uncertainty $\delta L_1$. One is the instrumental error due to the finite detector spatial resolution, $\delta l$. The other is the longitudinal straggling. Since the two uncertainties are not correlated, one has
\begin{equation}
\delta L_1  \leq \sqrt{2\delta l^2 + \delta L_{s}({\bf Q}_1)^2},
\label{dq1:1}
\end{equation}
where we use $\delta L_{s}({\bf Q}_1) = \max \left[ L_{sl}({\bf Q}_1), L_{st}({\bf Q}_1) \right]$ for the upper bound on the range straggling for the first recoil ion momentum ${\bf Q}_1$. There is no factor of 2 in the $\delta L_{sl}({\bf Q}_1)^2$ term because only the end point of the ion track is affected by straggling. 

From above, the magnitude error in ${\bf Q}_1$ is
\begin{equation}
\epsilon_r ({\bf Q}_1) \equiv \frac{\delta q_1}{q_1} = \frac{\delta L_1}{2E_1} \frac{dE_1}{dL_1},
\label{q1:er}
\end{equation}
which is accurate to the order $O(\beta_1^2) \sim 1\%$.

The direction error in ${\bf Q}_1$, which is the same as the error in ${\bf f}_1$, 
\begin{equation}
\epsilon_\phi({\bf Q}_1) = \epsilon_\phi({\bf f}_1) =\frac{\sqrt{4 \delta l^2 + 2\delta L_{s}({\bf Q}_1)^2 }}{L_1}.
\label{q1:ef}
\end{equation}

\begin{figure}[thbp] 
   \centering
   \includegraphics[width=3.0in]{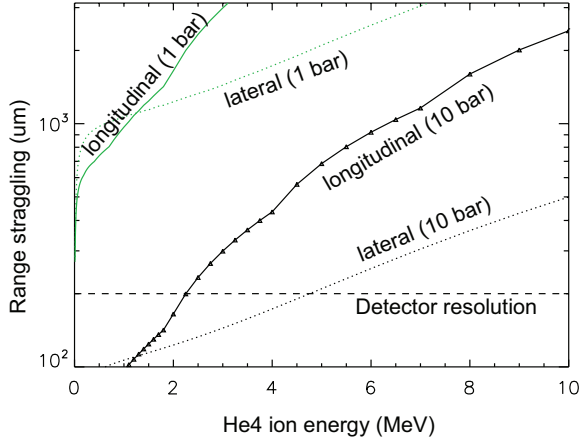} 
   \caption{Optimal detector pressure is to match the straggling range of an ion with the detector spatial resolution, or $\sqrt{2}\delta l \leq \delta L_{s}({\bf Q}_1)$ as in Eq.~(\ref{resolution:10x}).}
   \label{fig:He4Strag}
\end{figure}

From Eqs.~(\ref{dq1:1}) and (\ref{q1:ef}), there is no significant benefit to achieve an instrumental spatial resolution much below the range straggling, $\delta l \ll \delta L_{s}({\bf Q}_1)$. Matching the spatial resolution and ion range straggling $\delta l \sim \delta L_{s}({\bf Q}_1)$ is sufficient. Fig.~\ref{fig:He4Strag} shows that, by adjusting the stopping gas pressure, a detector resolution of 200 $\mu$m can be matched with the ion straggling range.

Next we examine the error in ${\bf P}_1$ as a function of ${\bf Q}_2$, ${\bf e}_1$ and ${\bf e}_2$.  It is convenient to replace the pair $\{${\bf e}$_1$, {\bf e}$_2 \}$ by a transformed pair $\{${\bf h}, $\zeta \equiv \sin \Theta_2 \}$ with the new unit vector {\bf h} defined as
\begin{equation}
{\bf h} = \frac{{\bf e}_1 - ({\bf e}_1\cdot {\bf e}_2) {\bf e}_2}{\zeta},
\end{equation}
and
\begin{equation}
 \zeta^2 = 1 - ({\bf e}_1\cdot {\bf e}_2)^2.
\end{equation}
$\Theta_2$, as defined in Fig.~\ref{fig:1}, is the angle between the vectors {\bf P}$_1$ (or {\bf e}$_1$) and {\bf P}$_2$ (or {\bf e}$_2$). In the new notation, Eq.~(\ref{p1:res1}) for $p_1$ becomes
\begin{equation}
p_1 =\frac{{\bf Q}_2 \cdot {\bf h}}{\zeta}.
\label{p1:22}
\end{equation}
We also have {\bf h}$\cdot${\bf e}$_1$ = $\zeta$ and {\bf h}$\cdot${\bf e}$_2$ =0. In general, the variations of $\delta {\bf Q}_2$, $\delta {\bf e}_1$ and $\delta {\bf e}_2$ are not independent of each other since they share a common point B. Due to the above length scaling, we use the following approximations
\begin{equation}
\epsilon_\phi({\bf P}_1) = \delta{\bf e}_1 = \delta {\bf e}_2 = 0.
\end{equation}
Therefore,
\begin{equation}
\begin{array}{ccl}
\delta p_1^2& =&  \displaystyle{ \frac{\delta {\bf Q}_2 \cdot {\bf h}}{\zeta} \frac{\delta {\bf Q}_2 \cdot {\bf h}}{\zeta}  } \\
 & \leq & \displaystyle{ \frac{ \delta {\bf Q}_2 \cdot \delta {\bf Q}_2}{\zeta^2}  } \\
 & = & \displaystyle{ \frac{ \delta q_2^2 + q^2_2 \delta {\bf f}_2 \cdot  \delta {\bf f}_2 }{\zeta^2}  }.
\end{array}
\end{equation}
As a result,
\begin{equation}
\begin{array}{ccl}
\epsilon_r({\bf P}_1) &=& \sqrt{\delta p_1^2}/p_1 \\
&=& \displaystyle{\frac{q_2}{p_1 \zeta} \sqrt{\epsilon_r^2({\bf Q}_2) + \epsilon_\phi^2({\bf Q}_2) } },
\end{array}
\label{p1:err}
\end{equation}
here $\epsilon_r({\bf Q}_2)$ and $ \epsilon_\phi({\bf Q}_2)$ are similar to Eqs. (\ref{q1:er}) and (\ref{q1:ef}), the error in ${\bf Q}_1$. 

At last,  we can calculate the error in ${\bf P}_0$ as a function of relative amplitudes and directions of ${\bf P}_1$ and ${\bf Q}_1$, that is, $\delta {\bf P}_0 =  \delta {\bf P}_1 + \delta {\bf Q}_1$. In terms of the components,
\begin{equation}
\delta p_0 {\bf e}_0 +   p_0 \delta {\bf e}_0 = \delta p_1 {\bf e}_1 +  \delta {\bf Q}_1,
\end{equation}
here again, we have neglected the error $\delta {\bf e}_1$. This equation can be broken down to the magnitude error
\begin{equation}
\begin{array}{ccl}
\delta p_0^2 &=& \delta p_1^2 ({\bf e}_1\cdot {\bf e}_0)^2 + ( \delta {\bf Q}_1\cdot {\bf e}_0)^2 \\
	&\leq& \delta p_1^2 \cos \Theta_1^2 + \delta {\bf Q}_1\cdot \delta {\bf Q}_1\\
	&=& \delta p_1^2 \cos \Theta_1^2 +  \delta q_1^2 + q_1^2 \delta {\bf f}_1 \cdot  \delta {\bf f}_1 ,
\end{array}
\label{p0:er}
\end{equation}
and the angular or direction error
\begin{equation}
\frac{\delta {\bf e}_0 \cdot \delta {\bf e}_0}{|\delta {\bf e}_0 |} = \displaystyle{\frac{1}{p_0}} (\delta p_1 \frac{{\bf e}_1\cdot \delta {\bf e}_0}{|\delta {\bf e}_0 |} + \frac{\delta {\bf Q}_1\cdot \delta {\bf e}_0}{|\delta {\bf e}_0 |}).
\end{equation}
The direction error can be simplified to
\begin{equation}
\delta {\bf e}_0 \cdot \delta {\bf e}_0 \leq \frac{1}{p_0^2}(\delta p_1^2 \sin \Theta_1^2 +  \delta q_1^2 + q_1^2 \delta {\bf f}_1 \cdot  \delta {\bf f}_1 )
\label{p0:ef}
\end{equation}
From Eq.~(\ref{p0:er}), one has
\begin{equation}
\epsilon_r({\bf P}_0) \leq \frac{1}{p_0}\sqrt{p_1^2 \epsilon_r({\bf P}_1)^2 \cos^2 \Theta_1+ q_1^2 \epsilon_r({\bf Q}_1)^2  + q_1^2 \epsilon_\phi({\bf Q}_1)^2}.
\label{p0:er2}
\end{equation}
From Eq.~(\ref{p0:ef}),
\begin{equation}
\epsilon_\phi({\bf P}_0) \leq \frac{1}{p_0}\sqrt{p_1^2 \epsilon_r({\bf P}_1)^2 \sin^2 \Theta_1+ q_1^2 \epsilon_r({\bf Q}_1)^2  + q_1^2 \epsilon_\phi({\bf Q}_1)^2}
\label{p0:ef2}
\end{equation}

The discussion thus far is on the five-point method. For the three-point method when the neutron TOF ($\tau_1$) between A and B in Fig.~\ref{fig:1} is available, $p_1$, instead of  Eq.(\ref{p1:res1}) or Eq.(\ref{p1:22}), is now given by
\begin{equation}
p_1 = \frac{L_{AB}}{\tau_1c}\displaystyle{ \frac{m_0 c}{\sqrt{1-(\frac{L_{AB}}{\tau_1c})^2}}}.
\end{equation}
The second fraction is the relativistic correction, which is on the order of $\tilde{\beta}_{1}^2$ with $\tilde{\beta}_{1}$ being the ratio of the neutron velocity to the speed of light. For a 10 MeV neutron, the relativistic correction is about 1\%. Therefore, with 1\% precision, it is sufficient to use the approximation
\begin{equation}
p_1 =  \frac{L_{AB}}{\tau_1c} \left[ 1 + \frac{1}{2}\left(\frac{L_{AB}}{\tau_1c}\right)^2 \right].
\end{equation}
Therefore,
\begin{equation}
\epsilon_r({\bf P}_1) = \sqrt{\left(\frac{\delta L_{AB}}{L_{AB}}\right)^2 +\left(\frac{\delta \tau_1}{\tau_1}\right)^2},
\label{tof:e1}
\end{equation}
As in the five-point method, we neglect the errors on the order of $\delta L_{AB}/L_{AB} \sim \delta l /\lambda < O(1\%)$, then
\begin{equation}
\epsilon_r({\bf P}_1) =\frac{\delta \tau_1}{\tau_1}.
\label{tof:e2}
\end{equation}
The component errors in ${\bf P}_0$ are still expressed by Eqs.~(\ref{p0:er2}) and (\ref{p0:ef2}), with the understanding that $\epsilon_r({\bf P}_1)$ is now given by Eq.~(\ref{tof:e2}). One may also try to use the TOF method to measure the ion momenta and obtain an error in the form given by Eq.~(\ref{tof:e1}). One now notices that the error due to first term $\delta L/L$ in Eq.~(\ref{tof:e1}) could be comparable or larger than the error due to the time measurement. Therefore, due to ion range straggling, the TOF measurement does not significantly reduce the errors in measurement of ${\bf Q}_i$.

\subsection{A neutron track example}
To obtain a quantitative understanding of the errors described by Eqs. (\ref{p0:er2}) and (\ref{p0:ef2}), we need the details of fast neutron scattering. The results are well known, in particular, when the angular anisotropy of the scattering in the Center-of-Mass (CM) frame, due to nuclear spin and other internal degrees of freedom, are neglected~\cite{BC36,Fermi49,Marshak47,GE:52,MW68}. The ratio of the neutron energies before and after the $n$th scattering is given by
\begin{equation}
\frac{\tilde{E}_n}{\tilde{E}_{n-1}} = \frac{A^2+1+2A\cos \theta_{n}}{(A+1)^2} = 1-\frac{\alpha}{2} (1- \cos \theta_{n}),
\label{En:1}
\end{equation}
and the scattering angle is given by
\begin{equation}
\cos \Theta_n = \frac{A \cos \theta_n +1}{\sqrt{A^2+2A\cos \theta_n +1}}.
\label{theta:rel1}
\end{equation}
Here following~\cite{BC36,WW:58}, the $\alpha $ parameter is defined as $\alpha \equiv 4A/(1+A)^2$. This definition of $\alpha$ is somewhat arbitrary~\cite{GE:52}. $\theta_n$ is the counterpart of $\Theta_n$ in the CM frame. While it is tempting to average Eqs.~(\ref{En:1}) and (\ref{theta:rel1}) and use them to calculate the errors described by Eqs. (\ref{p0:er2}) and (\ref{p0:ef2}). We do not use this approach here. A more rigorous treatment of angularly averaged errors, like Eq.~(\ref{p0:er2}), will require integrations over scattering angles $\Theta_1$ and $\Theta_2$, or equivalently $\theta_i$'s in the CM frame~\cite{Fermi49,MW68}. The angular average of a term like $\langle p_1^2 \epsilon_r({\bf P}_1)^2 \cos^2 \Theta_1 \rangle$ in Eq.~(\ref{p0:er2}) is not the same as the product of the averages of its factors, $\langle p_1^2\rangle \langle \epsilon_r({\bf P}_1)^2 \rangle \langle \cos^2 \Theta_1 \rangle$. Individual neutron tracks, however, do not depend on ensemble-averaged properties.

The neutron track example here satisfies
\begin{equation}
\frac{\tilde{E}_n }{\tilde{E}_{n-1}}  = 1-\frac{\alpha}{2},
\label{eq:avg1}
\end{equation}
which has nevertheless the same energy ratio as the average derived from Eq.~(\ref{En:1}). This result is equivalent to pick $\theta_n = \pi/2$. Correspondingly,
\begin{equation}
\cos \Theta_n  = \frac{1}{\sqrt{A^2+1}},
\label{theta:sel1}
\end{equation}
which is different from 2/(3A) when averaging Eq.~(\ref{theta:rel1}).


\subsection{Tracking errors}
\begin{figure}[thbp] 
   \centering
   \includegraphics[width=3.0in]{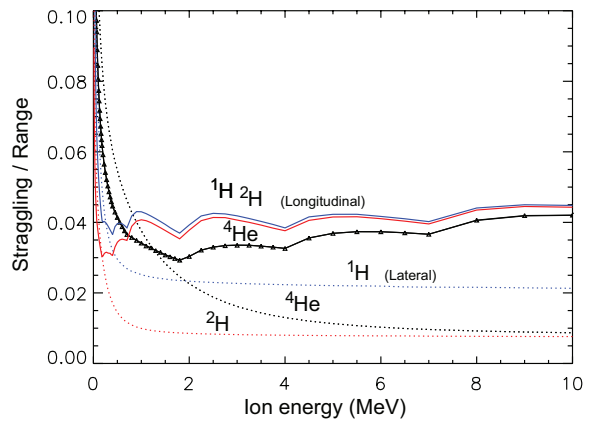} 
   \caption{Ratio of the range straggling to the total range at different ion energies. The solid lines are for longitudinal ranges. The dotted lines for lateral ranges. $^1$H is in methane gas. $^2$H is in deuterium gas. The results are not sensitive to pressure.}
   \label{fig:SRR}
\end{figure}

We now estimate the measurement errors of the neutron track that satisfies Eqs.~(\ref{eq:avg1}) and (\ref{theta:sel1}). 
Eq.~(\ref{q1:er}) gives
\begin{equation}
\begin{array}{ccl}
\epsilon_r ({\bf Q}_1) &=& \displaystyle{ \frac{\sqrt{2} \delta L_s({\bf Q}_1)}{2E_1} \frac{dE_1}{dL_1} } \\
&=&\displaystyle{ \frac{\sqrt{2}}{2} \frac{\delta L_s({\bf Q}_1)}{L_1} \frac{L_1}{E_1} \frac{dE_1}{dL_1}}\\
&\leq& \displaystyle{ \frac{\sqrt{2}}{2} \frac{\delta L_s({\bf Q}_1)}{L_1}. }
\end{array}
\label{q1M:er}
\end{equation}
To obtain Eq.~(\ref{q1M:er}), we use the approximation $(L_1/E_1) (dE_1/dL_1) \leq 1$, which is valid for $E_1$ above the Bragg peak of energy loss dE/dL. Based on SRIM calculations for $^1$H in Methane gas~\cite{SRIM}, the Bragg peak is at $E_1 =$ 0.06 MeV. For $^2$H in $^2$H gas, the Bragg peak is at $E_1$= 0.11 MeV. For $^4$He, the Bragg peak is at $E_1 =$ 0.65 MeV.
Eq.~(\ref{q1:ef}) gives
\begin{equation}
\epsilon_\phi({\bf Q}_1) =\frac{ 2\delta L_{s}({\bf Q}_1)}{L_1}.
\label{q1M:ef}
\end{equation}
Here we have chosen
\begin{equation}
\sqrt{2}\delta l \leq \delta L_{s}({\bf Q}_1).
\label{resolution:10x}
\end{equation}
The `$\leq$' sign is because two ions are involved. Therefore, $\sqrt{2}\delta l \sim \min_i (\delta L_{s}({\bf Q}_i))$. The results are shown in Fig.~\ref{fig:SRR} for a variety of conditions. The ratios are close to 4\% for longitudinal straggling, and less for lateral straggling. The results lead to $\epsilon_r ({\bf Q}_1)$ of about 3\% and $\epsilon_\phi({\bf Q}_1) $ of about 8\%. The results also apply to the second recoil ion, that is, $\epsilon_r ({\bf Q}_2) \sim$ 3\%  and $\epsilon_\phi({\bf Q}_2) \sim$ 8\%.
The error in {\bf P}$_1$, as given by Eq.~(\ref{p1:err}), is
\begin{equation}
\epsilon_r({\bf P}_1) =  \frac{\sqrt{2(A^2+1)}}{A+1}\sqrt{\epsilon_r^2({\bf Q}_2) + \epsilon_\phi^2({\bf Q}_2) }.
\end{equation}
Therefore, for $^1$H recoils, $\epsilon_r({\bf P}_1)$ = 8.5\%. For $^2$H recoils, $\epsilon_r({\bf P}_1)$ = 9\%. For $^4$H recoils, $\epsilon_r({\bf P}_1)$ = 10\%.
The errors in {\bf P}$_0$, as given by Eqs. (\ref{p0:er2}) and (\ref{p0:ef2}), are
\begin{equation}
\epsilon_r({\bf P}_0) \leq \frac{\sqrt{2A^2+1}}{A+1}\sqrt{\epsilon_r^2({\bf Q}_2) + \epsilon_\phi^2({\bf Q}_2) },
\label{p0:er4}
\end{equation}
and 
\begin{equation}
\epsilon_\phi({\bf P}_0) \leq  \frac{\sqrt{3} A}{A+1}\sqrt{\epsilon_r^2({\bf Q}_2) + \epsilon_\phi^2({\bf Q}_2) }.
\label{p0:ef4}
\end{equation}
For $^1$H recoils, $\epsilon_r({\bf P}_0) = \epsilon_\phi({\bf P}_0)$ = 7.4\%. For $^2$H recoils, $\epsilon_r({\bf P}_0)$ = 8.5\%, $\epsilon_\phi({\bf P}_0)$ = 10\%. For $^4$He recoils, $\epsilon_r({\bf P}_0)$ = 10\% and $\epsilon_\phi({\bf P}_0)$ = 12\%.

If TOF is used, $\epsilon_r({\bf P}_1)$ is free of errors from ion straggling. If $\epsilon_r({\bf P}_1)$ is small compared with ion straggling, then
\begin{equation}
\epsilon_r({\bf P}_0) = \epsilon_\phi({\bf P}_0) \leq \frac{\sqrt{2} A}{A+1}\sqrt{\epsilon_r^2({\bf Q}_2) + \epsilon_\phi^2({\bf Q}_2) }.
\end{equation}
For $^1$H recoils, $\epsilon_r({\bf P}_0) = \epsilon_\phi({\bf P}_0)$ = 6\%. For $^2$H recoils, $\epsilon_r({\bf P}_0) = \epsilon_\phi({\bf P}_0)$ = 8\%. For $^4$He recoils, $\epsilon_r({\bf P}_0) = \epsilon_\phi({\bf P}_0)$ = 10\%. These results are not significantly different from above. Therefore, a TOF accuracy 
\begin{equation}
\frac{\delta \tau_1}{\tau_1} \leq \sqrt{\epsilon_r^2({\bf Q}_2) + \epsilon_\phi^2({\bf Q}_2) } = 8\%
\label{tof:e4}
\end{equation}
is sufficient.


\subsection{Tracker dynamic range and size}
\begin{figure}[thbp] 
   \centering
   \includegraphics[width=3in]{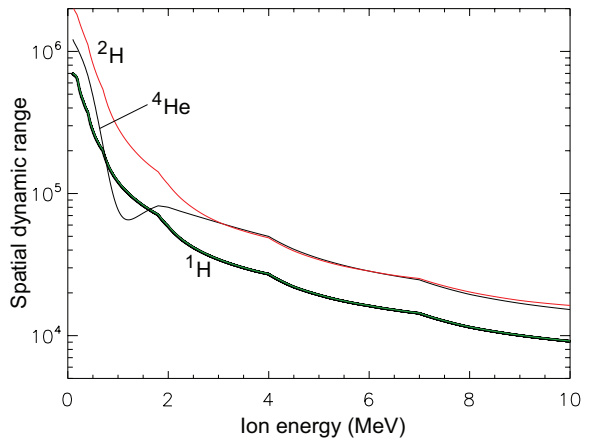} 
   \caption{Required spatial dynamic range $\mathfrak{D}$ as a function of ion energy.  Here $\delta l = \delta L_s /\sqrt{2}$ with $\delta L_s$ being longitudinal range straggling. $^1$H is in the methane gas. $^2$H is in the deuterium gas. The results are not sensitive to pressure.}
   \label{fig:2}
\end{figure}

The spatial dynamic range ($\mathfrak{D}$) of a fast-neutron tracker may be defined as
\begin{equation}
\mathfrak{D} \equiv \frac{\lambda}{\delta l}, 
\end{equation}
the ratio of the neutron mean free path to the spatial resolution of the track recording instrument. From above, $\delta l = \delta L_s /\sqrt{2}$, the spatial resolution is chosen to be comparable to the ion range straggling. The spatial dynamic range is plotted as a function of ion energy in Fig.~\ref{fig:2} for three different ions. $\mathfrak{D}$ is in the range of 10$^4$ to a few times 10$^5$  for 1 to 10 MeV ions, independent of the density of the detector medium.

When multiple collisions are taken into account, the size of the neutron tracker for efficient detection to accommodate $n$ collisions can be estimated as the following.
\begin{equation}
{\bf R}(n) =  \sum_{j=0}^{n-1}  {\bf s}_j ,
\end{equation} 
with ${\bf R}(n)$ being the total range and ${\bf s}_j$ the neutron range prior to the $(j+1)$th collision. Following~\cite{Fermi2}, 
\begin{equation}
\langle {\bf s}_j \rangle = \langle {\bf s}_0 \rangle  \langle  \cos \Theta_1 \rangle^j,
\end{equation}
with $\langle \cos \Theta_1 \rangle =2/(3A)$. Therefore
\begin{equation}
\langle {\bf R}(n) \rangle = \langle {\bf s}_0 \rangle \sum_{j=0}^{n-1}  \left( \frac{2}{3A}\right)^j.
\end{equation} 
Drop the vector notation and use $S(n)$ for the average detector size for $n$ collisions, from the magnitude of $\langle {\bf s}_0 \rangle = \lambda_0$, we found
\begin{equation}
S(n) \le  S (\infty) =  \displaystyle{ \frac{\lambda_0}{1-\frac{2}{3A}} }.
\end{equation}
For $^1$H, S(2) = 5$\lambda_0$/3, S(3) = 19$\lambda_0$/9. For $^2$H, S(2) = 4$\lambda_0$/3, S(3) = 13$\lambda_0$/9. For $^4$He, S(2) = 7$\lambda_0$/6, S(3) = 43$\lambda_0$/36. The deviation from the average detector size can be defined as
\begin{equation}
\sigma_S^2 (n) \equiv  \langle {\bf R}^2(n) \rangle - S^2(n).
\end{equation}
We are only interested in $n \leq 3$, which can be calculated readily.
\begin{equation}
\sigma_S^2 (1) = \lambda^2_0,
\end{equation}
\begin{equation}
\sigma_S^2 (2) = \lambda^2_0 \left( 3 - \frac{4}{9A^2} \right)  ,
\end{equation}
and
\begin{equation}
\sigma_S^2 (3) = \lambda^2_0 \left( 5 + \frac{4}{3A} - \frac{4}{9A^2} -\frac{16}{27A^3} -  \frac{16}{81A^4}\right).
\end{equation}


\section{Track measurement}
\label{sec:TM}

There are at least two types of ion trackers that are useful to real-time tracking of fast neutrons. One is based on charge collection, as in various multi-wire proportional chambers (MWPCs)~\cite{Charpak:1968,Sauli:1977,Sauli:1999}. The other is based on photon collection, as in optical imaging.  The purposes are the same: a.) to pinpoint the positions of the neutron elastic collisions or locate the vertices A, B, and C in Fig.~\ref{fig:1} and b.) to find the lengths and the directions of the recoil ion tracks, {\bf Q}$_1$ and {\bf Q}$_2$. The number of charged pairs and photons produced during the stopping of the recoil ions determines the signals available. Both the charge and photon methods require single-particle sensitivity, that is, sensitivity to a single electron or a single photon. 

\subsection{Charge-based ion tracking}

MWPCs deliver superb charge sensitivity and spatial resolution~\cite{Charpak:1968,Sauli:1977,Sauli:1999}. The latest high gain operations can achieve a single-stage charge multiplication by a factor of 10$^3$ to 10$^6$, sufficient for single charge detection using low-noise preamplifiers. Fine spatial resolutions less than 100 $\mu$m are possible with micro-pattern readouts, such as micromegas and gas electron multipliers (GEMs)~\cite{Sauli:1999}. 

In the absence of a magnetic field, the spatial resolution in an MWPC ($\sigma_d$) is limited by the diffusion of primary electrons~\cite{BR:2008},
\begin{equation}
\sigma_d= \sqrt{\frac{4\varepsilon_e\mathfrak{L}}{3eE_b}},
\label{dft:2}
\end{equation}
here $E_b$ is the bias electric field strength in the drift region, $\mathfrak{L}$ is the drift distance, $\varepsilon_e$ is the so-called characteristic energy of drift electrons. Adding a magnetic field may reduce diffusion across the field lines as in time projection chambers~\cite{Nygren:1978,Hilke2010}. However, besides the overburden of bulky coils or heavy magnets, the benefits of a magnetic field are further compromised here since the uncertainties of ion tracking also depend on the position accuracy along the magnetic field, which is not improved by the magnetic field.  In the thermal limits, $\varepsilon_e = 3/2 k_BT_e$. In another limit, when $\varepsilon_e = 3 g_d \lambda e E_b/4 \gg k_BT_e$, one has $\sigma_d = \sqrt{g_d \lambda \mathfrak{L}}$ with a coefficient $g_d$ on the order of one. Typical resolution ranges from less than 100 $\mu$m to several hundred $\mu$m, depending on the drift distance $\mathfrak{L}$. The maximum drift distances for different 1 MeV recoil ions have been listed in Table.~\ref{tb:1}. It is found that the drift distances are on the order of 10 cm under the standard temperature and a pressure at 10 bar. Charge diffusion will make it impractical to recover ion tracks for longer drift distances. 

The spatial dynamical range $\mathfrak{D}$ is about 500 for a 10-cm 10-bar detector cell. A large cubic tracker with at least 20 cells each side would be required to reach a $\mathfrak{D}$ of 10$^4$.  Close to 100\% intrinsic neutron tracking efficiency requires 20$^3 \sim$ 10$^4$ cells. For fewer cells, the detection efficiency decreases and is proportional to the number of cells.  A detection efficiency of 0.1\% will only need tens of detector cells. For a detector with a few cells, the intrinsic efficiency is in the range of 10$^{-6}$ to 10$^{-4}$. In this case, to pinpoint a neutron source requires an increase in detection time to compensate for the reduced intrinsic efficiency.


\begin{table}[hb]
\caption{\label{tab:table1}%
Estimates of the maximum drift distances before the initial charge cloud becomes too diffusive. No magnetic field is assumed. The number of the elemental charge after drift is 10 in a square pixel with a size $l_p$. $X$ is the recoil ion range at an energy of 1 MeV. $\delta L_s$ is the range straggling. W is the energy required to produce each charge pair. All gases are assumed to be at 10 bar STP.
}
\centering
\begin{tabular}{lccc}
\hline
\textrm{Recoil ion}\footnote{$^1$H in the CH$_4$ gas.}&
\textrm{$^1$H}&
\textrm{$^2$H}&
\textrm{$^4$He}\\\hline
X (mm) & 2.36 &  5.88 & 2.84 \\
$\delta L_s$ (mm) & 0.10 &  0.24 & 0.10 \\
$V_b$ (V)\footnote{$V_b=E_b\mathfrak{L}$.} & 1000 & 1000 & 1000\\
W (eV)\footnote{Data from \cite{BR:2008}. W's for $^2$H and $^1$H are assumed to be the same.} & 29.1 & 36.4 & 46.0 \\
$l_p$ ($\mu$m) & 200 & 200 & 200\\
$\mathfrak{L}$ (cm) & 10 & 9.1 & 8.1\\
\hline
\end{tabular}
\label{tb:1}
\end{table}

An MWPC normally only measures the projection of a 3D ion track onto a 2D plane. To obtain the third coordinate along the drift direction, a time projection technique may be used. Time projection measurement based on the light emitted from the ion tracks will be discussed in Sec.~\ref{subsec:time}.

\subsection{Light-field imaging}
Ion tracking based on light emitted along the ion tracks has yet to be demonstrated~\cite{Roccaro:2009}. Besides electron-ion pair creation along its track, a recoil ion can also induce light emissions by exciting electrons from ground states. De-excitations emit light. Another type of light emission is due to energetic electrons produced by the ion. Head-on collisions between the ion and electrons produce the most energetic electrons called $\delta$-rays by J. J. Thomson. Lower energy electrons are ejected when the ion and electrons interact farther away. Except for $\delta$-ray-induced light emissions, which can be mm or farther away from an ion track, most light emissions come from lower energy electrons, which closely follow the ion track and can be used for ion tracking. Optical methods remove track blur due to charge diffusion.

The problems, sharing the same flavors as in charge-based methods, are to determine the positions and lengths of the light-emitting columns due to ion stopping. Although the fine spatial structures of the light columns smaller than ion range straggling is not essential, the limited number of available photons is a challenge. In-situ photon multiplication is not considered. We also neglect light scattering and absorption before reaching the detectors. 

Light-field imaging (LFI) extracts scene information from different properties of a photon, its direction, wavelength, arrival time and polarization~\cite{Adelson:1992,AW:1992}.  LFI can minimize the number of photons required to localize a photon-starved scene. As few as two crossing views are sufficient to locate a point source~\cite{Longuet:1981,Weng:1989,HartZiss:2004}. Any additional detail from the scene, such as the source structure, requires more photons~\cite{Rose:1976}. Photon energies or wavelengths dictate detector material use and construction for high detection efficiency. Photon arrivals at the detectors are randomly delayed from the times of excitation. The random photon emissions follow the Poisson distribution with time constants equal to the inverse transition probabilities, or the half lives of the excited states.

\begin{figure}[thbp] 
   \centering
   \includegraphics[width=2.5in]{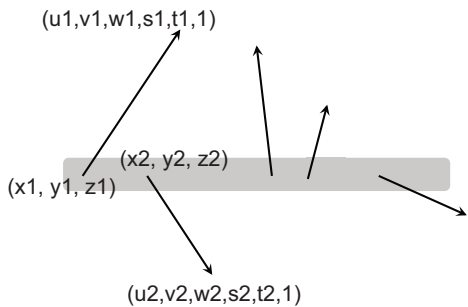} 
   \caption{An ion track is approximated by a line segment (shaded). Every photon emitted, represented by a ray, has to anchor in the ion track. In general, the rays are not coplanar.}
   \label{fig:IonLFI1}
\end{figure}

We may estimate ideally, how many photons are needed to determine an ion track, which is approximated by a line segment, shown as the shaded region in Fig.~\ref{fig:IonLFI1}. The question is to locate the two end-points. Equivalently,  one may also locate an end point and measure the track length and the direction, ($x_0$, $y_0$, $z_0$, $s_0$, $t_0$, 1). Here the first three parameters are the coordinates of a point in the ion track. ($s_0$, $t_0$, 1) is the direction of the ion track. Each photon, represented by a ray as ($u_1$, $v_1$, $w_1$, $s_1$, $t_1$, 1), has to have an anchor point ($x_1$, $y_1$, $z_1$) in the ion track. Each photon ray therefore supplies four independent equations that describe the fact that the anchor point has to be on the photon ray and in the ion track simultaneously, Eqs.~(\ref{ray1:1}) and (\ref{ray1:2}), in which the unknowns are ($x_0$, $y_0$, $z_0$, $s_0$, $t_0$) and ($x_1$, $y_1$, $z_1$).

\begin{equation}
\frac{x_1-x_0}{s_0} = \frac{y_1-y_0}{t_0} = z_1-z_0,
\label{ray1:1}
\end{equation}
and
\begin{equation}
\frac{x_1-u_1}{s_1} = \frac{y_1-v_1}{t_1} = z_1-w_1.
\label{ray1:2}
\end{equation}
Another four photon rays, or for a total of five photon rays, would have a sufficient number of equations to solve for a total of five unknown anchor points and the unknowns for the ion track. For the track length, one needs to trace an additional photon.

Although less than ten photons are sufficient for ion tracking, it will take more photons to locate the end-points of the ion track and measure the ion track length more accurately. The least number of photons needed is determined by Poisson statistics. For a signal-to-noise of five, the minimum number of photons needed per resolvable segment of the ion track is 25. If the whole ion track is considered, the total number of photons is to multiply 25 by the number of resolvable segments, which is the ratio of the ion track length to straggling length, or 25 From Fig.~\ref{fig:SRR}. Therefore the total number of photons is at least 625 for MeV recoil ion tracks.

Photons emitted along an ion track come from two sources of electrons. Multiply scattered electrons (MSEs)~\cite{GS40,Moliere48,Bethe53,mm65} and $\delta$-rays. MSEs have lower energies than $\delta$-rays and form the primary track~\cite{MM:1962,OL:1967}. The relative numbers of MSEs to $\delta$ rays is proportional to the ratio of multiple scattering cross section to two-body collision cross section. The ratio may be characterized by a $\Lambda$ parameter~\cite{Chen06,BS03},
\begin{equation}
\Lambda \equiv \frac{\lambda_D}{b_0}.
\end{equation}
Here, the so-called shielding length ($\lambda_D$) is given by
\begin{equation}
\lambda_D=\sqrt{\frac{\epsilon_0 I_0}{n_0 Z e^2}}
\end{equation}
with $I_0$ being the ionization potential of the stopping medium, $n_0$ the atomic density of the medium, $Z$ the ionic charge and $\epsilon_0$ the vacuum permittivity. The distance of the closest encounter $b_0$ is given by
\begin{equation}
b_0 = \frac{Ze^2}{4\pi \epsilon_0 E_\delta^{\max}}.
\end{equation}
The maximum energy of a $\delta$-ray is given by $E_{\delta}^{\max} = 4(m_e/M_p) (E_i/A_i)$, which is the largest for protons with $A_i=1$. At the present, $E_i \lesssim 10$ MeV.  Therefore, $E_{\delta}^{\max}  \lesssim 22$ keV. Both ranges of MSEs and $\delta$ rays may be calculated from the continuous slowing-down approximation (CSDA). The lateral distances of $\delta$ ray photons from the ion trajectory is larger than MSE photons. In the hydrogen gas, the CSDA range for a 22 keV electron is 4.54 $\times 10^{-4}$ g/cm$^2$~\cite{ESTAR}, or about 5 mm at 10 bar pressure and room temperature.

The primary track has a sharp radius, which is determined by MSEs with an average energy up to 1 to 4 keV~\cite{MM:1962}. The primary column radius, the same as the average MSE range, may be approximated by $R_e = 9.93 \times 10^{-6} E_e$ [g$\cdot$cm$^{-2}$keV$^{-1}$], independent of stopping medium~\cite{KK:1968}.  

The number of photons per unit length along an ion track is given by
\begin{equation}
\frac{dN_\lambda}{dL} = - (1-F) \frac{1}{W}\frac{dE}{dL}\frac{f \langle \sigma_l v_e \rangle }{\langle \sigma_i v_e \rangle }
\label{ph:1}
\end{equation}
at a wavelength $\lambda$. $W$ is again the average energy for ion-pair production, which is assumed to be the same for both MSEs and $\delta$-rays. $\langle \sigma_l v_e \rangle /\langle \sigma_i v_e \rangle$ is the ratio of the averaged excitation rate to an atomic level $l$ to the ionization rate. $f \leq 1$ is the branching ratio for $\lambda$. $F$ is the fraction of energy that goes to the $\delta$-ray production and does not contribute to the light production in the primary track. $F$ was found to depend on $E_i/A_i$ only~\cite{MM:1962}. After an ion energy falls below a certain threshold, $\delta$-rays can no longer be produced, and $F = 0$. From $E_{\delta}^{\max} = 4(m_e/M_p) (E_i/A_i) \leq 1$ keV,   ones finds $E_i = $ 0.46 MeV for protons, 0.92 MeV for deuterons and 1.8 MeV for $^4$He's. These numbers are small compared with the values in Fig. 3 of~\cite{MM:1962}. If one uses the threshold of 4 keV instead, then $E_i = $ 1.8 MeV for protons, 3.6 MeV for deuterons and 7.2 MeV for $^4$He's.  Here we use $F = 0.25$ for simplicity, which may underestimate the photon production by up to 25\%. Furthermore, Eq.~(\ref{ph:1}) does not include photons from ion-pair recombination. 

\begin{figure}[thbp] 
   \centering
   \includegraphics[width=3.0in]{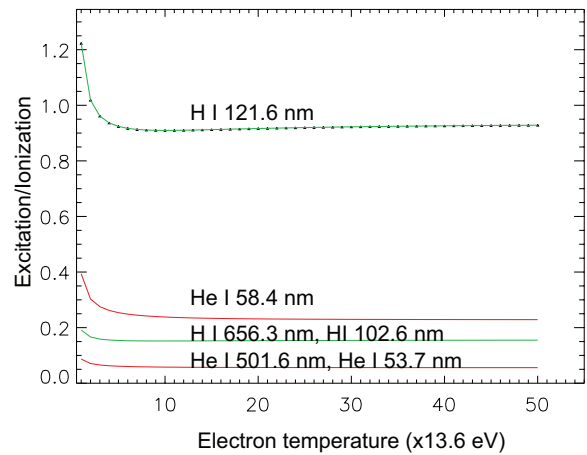} 
   \caption{Calculated ratio $\langle \sigma_l v_e \rangle /\langle \sigma_i v_e \rangle$ for a gaussian energy distribution of electrons in the primary track. The results are insensitive to the electron temperature. We use 0.91 for H I 121.6 nm, 0.24 for He I 58.4 nm, 0.15 for H I 656.3 nm and H I 102.6 nm, 0.058 for He I 501.6 nm and He I 53.7 nm. }
   \label{fig:pr1}
\end{figure}

Based on the references~\cite{NIST2,Wiese:1966,asd:2012}, the branching ratios are calculated for several wavelengths of $^1$H, $^2$H and $^4$He, Table.~\ref{tab:table2}. The spectral properties of $^2$H and $^1$H are treated the same and not repeated. The electron energy distribution in the primary track is assumed to be a gaussian with an adjustable temperature parameter. The resulting ratios $\langle \sigma_l v_e \rangle /\langle \sigma_i v_e \rangle$  are shown to be insensitive to the electron temperature, Fig.~\ref{fig:pr1}.

\begin{table}[htb]
\caption{\label{tab:table2}%
Calculated branching ratios ($f$) for several wavelengths of the $^1$H and $^4$He gas. The largest numbers of photons are expected at these wavelengths. The results for the $^2$H gas are expected to be the same as those of $^1$H and not listed.
}
\centering
\vspace{0.5cm}
\begin{tabular}{lccr}
\hline
\textrm{Wavelength (nm)}&
\textrm{Transition}&
\textrm{$A_k$(10$^8$ s$^{-1}$})\footnote{Transition probability.} &
\textrm{$f$}\footnote{Collisional depopulation rate is neglected.}\\\hline
H I 121.6 ($L_\alpha$)&  2p $^2P_{3/2,1/2} \rightarrow$ 1s $^2S_{1/2}$& 4.70 &1.0  \\
H I  656.3 ($H_\alpha$)&  3p $^2P_{3/2,1/2} \rightarrow$ 2s $^2S_{1/2}$& 0.441 &0.44  \\
H I  102.6 ($L_\beta$)&  3p $^2P_{3/2,1/2} \rightarrow$ 1s $^2S_{1/2}$&  0.558 &0.56  \\
He I 58.4 &  2p1s $^1P_1 \rightarrow$ 1s$^2$ $^1S_0$ & 18.0 &1.0  \\
He I 501.6 & 3p1s $^1P_1 \rightarrow$ 2s1s $^1S_0$ & 0.134 &0.023 \\
He I 53.7 &  3p1s $^1P_1 \rightarrow$ 1s$^2$ $^1S_0$ & 5.66 & 0.98 \\
\hline
\end{tabular}
\end{table}

\begin{figure}[thbp] 
   \centering
   \includegraphics[width=3.0in]{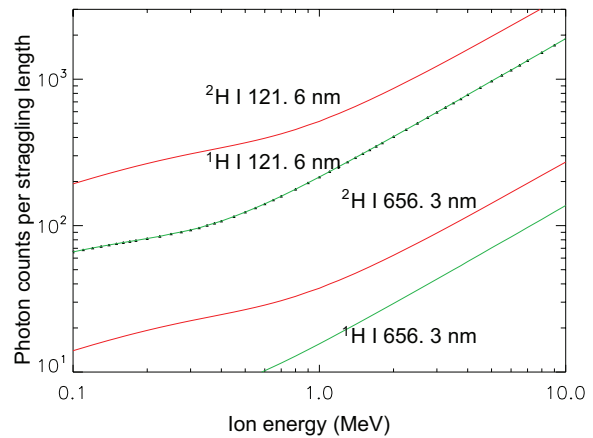} 
   \caption{Calculated photon counts per lateral straggling length in the primary track for $^1$H and $^2$H. $^1$H is in pure hydrogen gas. $^2$H is in deuterated methane. The results are not sensitive to the mass density (pressure). }
   \label{fig:pr5}
\end{figure}

\begin{figure}[thbp] 
   \centering
   \includegraphics[width=3.0in]{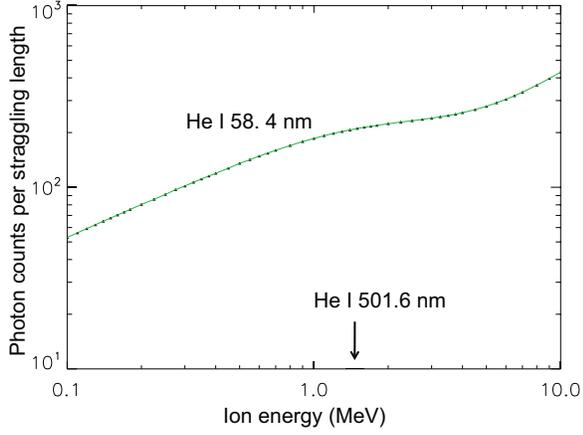} 
   \caption{Calculated photon counts per lateral straggling length in the primary track for $^4$He. $^4$He is in pure helium gas. The results are not sensitive to the mass density (pressure).  Only the He I 58.4 nm has sufficient photons above the threshold of 10 counts per straggling length. The visible line He I 501.6 nm has too few number of photons.}
   \label{fig:pr9}
\end{figure}

Combining the results from Table.~\ref{tab:table2} and Fig.~\ref{fig:pr1}, we obtain the expected photon densities along the primary tracks, Fig.~\ref{fig:pr5} for $^1$H ($^2$H) and Fig.~\ref{fig:pr9} for $^4$He. Spectral lines H I 121.6 nm, H I 656.3 nm and He I 58.4 nm have sufficient photons per straggling length for imaging. He I 501.6 nm is not a good candidate because of the paucity of photons. Using lense-based method to measure photon directions (see below), we will also run into problems with He I 58.4 nm, since the shortest transmission wavelength of any known material is about 105 nm using a LiF window or lense. It is worthwhile to look into lenseless methods~\cite{AlW:2009} to detect directions of VUV photons like He I 58.4 nm, but we do not elaborate here. 

\begin{figure}[thbp] 
   \centering
   \includegraphics[width=2.5in]{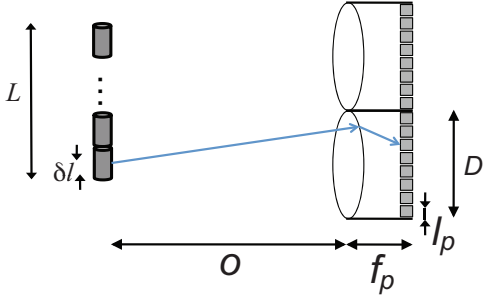} 
   \caption{An approach to photon-starved LFI is to use a lenslet array that surrounds the scene, so that close to 100\% photons are detected. Each lenslet sorts photons into different pixels of a detector according to their incoming directions. The directional uncertainty is diffraction limited. The possibility of super-resolution is not considered.}
   \label{fig:LC1}
\end{figure}

A simple 4$\pi$ detector without any optics can capture all the light but is not useful because it can not resolve the directions of photons. On the other hand, a detector, in conjunction with small pin-hole collimators, can meet the angular resolution requirements but has too small a photon collection efficiency to be useful. A possible method is to use a lenslet array that covers $\sim 4\pi$ solid angle so that most photons are captured. For each lenslet in the array, there is a two-dimensional detector array at the focal plane, Fig.~\ref{fig:LC1}. Each lenslet in the array sorts the incoming photons into different detector pixels according to photon directions. The pixel size of the detector array ($l_p$) matches the diffraction limited spatial resolution, $l_p = \alpha_r f_p$, with $f_p$ being the focal length and the angular resolution $\alpha_r = 1.22 \lambda /D $~\cite{Smith:2000,BW:1999}. Here $D$ is the diameter of the lenslet, as in Fig.~\ref{fig:LC1}. 

The spatial resolution at the scene is given by 
\begin{equation}
\delta l = \alpha_r O + D = 1.22\frac{\lambda O}{D}+D,
\end{equation} 
with $O$ being the distance between the ion track and a lenslet. The best resolution of $\delta l_{\rm min} = 2D$ is therefore reached when
\begin{equation}
D^2 = 1.22 \lambda O.
\end{equation}
For $\delta l_{\rm min} = \delta L_s / \sqrt{2}$, 
\begin{equation}
D = \frac{\delta L_s}{2\sqrt{2}}.
\end{equation}
For $\delta L_s$ = 200 $\mu$m, $D$ =71 $\mu$m. The distance $O$ is inversely proportional to the wavelength. At H I 121.6 nm, $O$ = 34 mm.  At H I 656.3 nm, $O$ = 6.2 mm. At He I 58.4 nm, $O$ = 70 mm. Correspondingly, the spatial dynamic range of a detector cell, $\mathfrak{D} = 2O/\delta L_s$, is  337, 62 and 702. On average, no more than one photon will reach each lenslet. 







\subsection{Time measurement}
\label{subsec:time}
Recording time information is useful and in some cases necessary in fast neutron tracking. As in Sec.~\ref{princ:1}, the number of required recoil collisions reduces from three to two if neutron TOF is available. Timing of neutron collisions can be used in charge-based ion tracking to obtain the ion drift time and the third coordinate along the drift electric field. In LFI, photon TOF from an ion track to the detectors is redundant because of Eqs.~(\ref{ray1:1}) and (\ref{ray1:2}).  We shall only discuss timing measurements based on light emissions. Other methods, such as sensing neutron magnetic momentum, may also be possible but are not discussed. 

\begin{figure}[thbp] 
   \centering
   \includegraphics[width=3.0in]{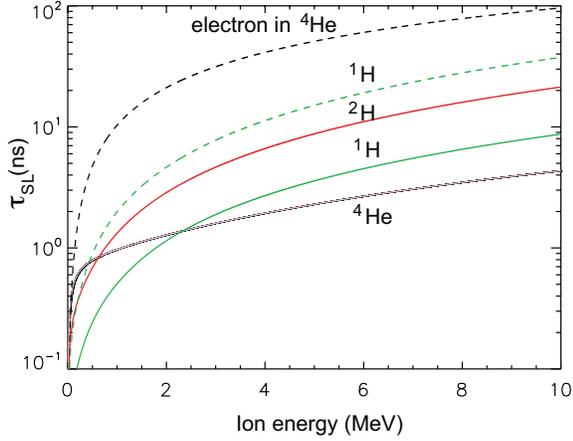} 
   \caption{Average ion slowing-down time in different gases as a function of the initial ion energy. From the top to the bottom are for electrons in  $^4$He, $^1$H in $^1$H gas, $^2$H in $^2$H gas, $^1$H in methane gas, and $^4$He in $^4$He gas. All gas pressures are at 10 bar STP.}
   \label{fig:3}
\end{figure}

For a photon emission ($\lambda$) with a transition probability $A_{l\lambda}$, the number of atoms in the corresponding excited state ($N_l$) as a function of time is given by
\begin{equation}
\frac{d N_l(t)}{dt} = -N_l(t) \sum_{\lambda'} A_{l\lambda'} - (1-F)  \frac{1}{W}\frac{dE}{dt}\frac{ \langle \sigma_lv_e \rangle }{\langle \sigma_i v_e \rangle }
\label{em:1}
\end{equation}
Summing over different wavelengths is due to the fact that several de-excitation transitions may be possible. The number of photons with the wavelength $\lambda$ as a function of time $N_\lambda(t)$ satisfies
\begin{equation}
\frac{d N_\lambda(t)}{dt} = A_{l\lambda} N_l(t).
\label{em:2}
\end{equation}
Here $t=0$ is when a neutron collides with the ion, which coincides with the start of ion motion. Combining Eqs.~(\ref{em:1}) and (\ref{em:2}),
\begin{equation}
\frac{dN_l(t)}{dt} = - \frac{1}{f} \frac{dN_\lambda(t)}{dt} - (1-F)  \frac{1}{W}\frac{dE}{dt}\frac{ \langle \sigma_lv_e \rangle }{\langle \sigma_i v_e \rangle }.
\label{em:3}
\end{equation}
Eq.~(\ref{em:3}) is consistent with Eq.~(\ref{ph:1}) by integration over time from the beginning of the ion motion until ion stopping or the end of de-excitation time. Solving Eq.~(\ref{em:1}) for $N_l(t)$, 
\begin{equation}
N_l(t) = - \int_0^t ds e^{\gamma_l (s-t)} (1-F)  \frac{1}{W} \frac{dE}{ds} \frac{ \langle \sigma_lv_e \rangle }{\langle \sigma_i v_e \rangle },
\label{sol:10}
\end{equation}
here we define $\gamma_l \equiv \sum_\lambda A_{l\lambda}$. Solving Eq.~(\ref{em:3}) for $N_\lambda(t)$, 
\begin{equation}
N_\lambda(t) = f (1-F)  \frac{E(0) - E(t)}{W} \frac{ \langle \sigma_lv_e \rangle }{\langle \sigma_i v_e \rangle } - f N_l(t).
\label{sol:11}
\end{equation}
The slowing-down times of recoil ions are comparable or longer than de-excitation times, which are around 1 ns, see Fig.~\ref{fig:3}. Here the slowing-down times ($\tau_{SL}$) are given by 
\begin{equation}
\tau_{SL} = \int_0^{E(0)} \frac{dE}{\beta c} \left( \frac{dE}{dL} \right)^{-1},
\end{equation}
where $dE/dL$ is the energy loss of the recoil ions per unit length, $\beta$ is the ratio of an ion speed to the speed of light. We used SRIM to calculate $dE/dL$ for different ions~\cite{SRIM} and ESTAR database maintained by NIST to calculate $dE/dL$ for fast electrons~\cite{ESTAR}. 

The time of neutron collision (t=0) can not be measured since there is no signal available.  The emission of a photon obeys Poisson statistics, so there is a theoretical possibility that at $t=0$ a photon with the wavelength $\lambda$ is emitted and is detected after a time delay which is the sum of photon TOF and detector response time. The average time that characterizes the Poisson statistics of photon emission $\bar{t}_0$ satisfies
\begin{equation}
N_\lambda(\bar{t}_0) = 1
\end{equation}
with $N_\lambda(t)$ given by Eq.~(\ref{sol:11}). Using the approximation $dE/dt = $const, which is justified {\it a posteriori} , 
\begin{equation}
\bar{t}_0 = \sqrt{\frac{2 \tau_{ex}}{ f\gamma_l}}
\end{equation}
with  $\tau_{ex}$ being defined as
\begin{equation}
\tau_{ex} \equiv \frac{W}{1-F} \frac{ \langle \sigma_iv_e \rangle }{\langle \sigma_l v_e \rangle } \left (\frac{dE}{dt}\right)^{-1}.
\end{equation}

\begin{figure}[thbp] 
   \centering
   \includegraphics[width=3.0in]{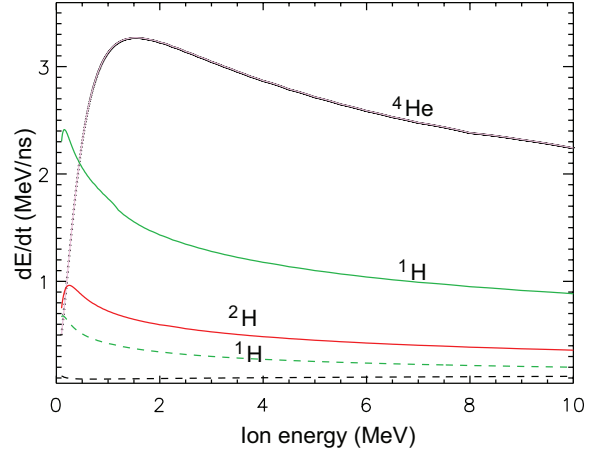} 
   \caption{Ion energy loss rate in different gases as a function of the initial ion energy. From the top to the bottom are for $^4$He in $^4$He gas, $^1$H in methane gas,  $^2$H in $^2$H gas,  and $^1$H in $^1$H gas. All gas pressures are at 10 bar STP.}
   \label{fig:4dedt}
\end{figure}
Using data from Fig.~\ref{fig:4dedt} and Table.~\ref{tab:table2}, we may calculate $\bar{t}_0$ for different scenarios. The results are given in Table.~\ref{tab:table3}.  
\begin{table}[htb]
\caption{\label{tab:table3}%
Calculated $\bar{t}_0$ for various scenarios. The ions are at 2 MeV.
}
\centering
\vspace{0.4cm}
\begin{tabular}{lccr}
\hline
\textrm{Wavelength (nm)}&
\textrm{$\gamma_l$ (ns$^{-1}$)}&
\textrm{$\tau_{ex}$(ps)} &
\textrm{$\bar{t}_0$ (ps)}\\\hline
H I 121.6 ($L_\alpha$)& 0.47 & 0.03 &11  \\
H I  656.3 ($H_\alpha$)& 0.10 & 0.18 & 92  \\
H I  102.6 ($L_\beta$)& 0.10  & 0.18  & 81  \\
D I 121.6 & 0.47 & 0.09 &20  \\
D I  656.3 & 0.10 & 0.54 & 157  \\
D I  102.6 & 0.10  & 0.54  & 139  \\
He I 58.4 & 1.8  & 0.79 & 9.4  \\
He I 501.6 & 0.58 & 0.33 & 222 \\
He I 53.7 & 0.58 & 0.33 & 34 \\\hline
\end{tabular}
\end{table}
To take advantage of the ps time response of the photons, efficient, single-photon sensitive and ps-resolution detector would be ideal. These detectors do exist and their performances are suitable for LFI~\cite{Cova:1996,Niclass:2006}. If 100 ps time resolution is indeed obtained, from Eq.~(\ref{tof:e4}), the neutron TOF for 8\% resolution will be 1.25 ns. At 10 MeV, the flight length required is about 5 cm.



\section{Discussion}
\label{sec:DC}
\begin{figure}[thbp] 
   \centering
   \includegraphics[width=3.0in]{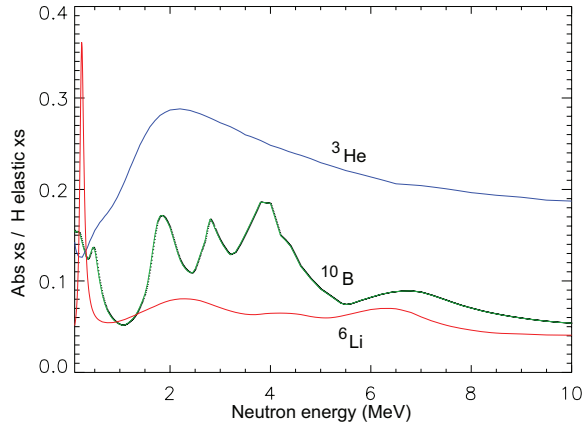} 
   \caption{Ratios of the neutron absorption cross sections of $^3$He, $^6$Li and $^{10}$B to the elastic cross section of $^1$H(n,n)$^1$H for neutrons in the MeV energy range. The data are from~\cite{t2:lanl}.}
   \label{fig:abs}
\end{figure}

Our analysis has mostly focused on elastic collisions of MeV neutrons with $^1$H, $^2$H and $^{4}$He, all of which are abundant and can exist in the gas phase. For $^1$H and $^2$H, abundant organic compounds also exist. There are a few low-Z elements that have relatively large absorption cross sections for MeV neutrons. In Fig.~\ref{fig:abs}, the ratios of the fast neutron absorption cross sections to the elastic cross section of $^1$H(n,n)$^1$H are shown for $^3$He, $^6$Li and $^{10}$B. $^3$He is the most attractive in that it has the largest absorption cross section and the smallest Q-value at 0.76 MeV. However, due to the short supply of $^3$He, a large volume of $^3$He may not be available. In a small volume, with the detector areal density ($nL$) satisfying $\sigma  n L \ll 1$ and $\sigma$ being either the neutron absorption or elastic collision cross section,  the efficiency of an absorptive neutron tracker can be higher than that of an elastic tracker. Since an absorptive tracker only requires one collision, the intrinsic efficiency is given by $\epsilon_1 \sim \sigma_{abs} n L $. For an elastic tracker that requires three collisions, the intrinsic efficiency $\epsilon_3 \sim (\sigma_{el} n L)^3$. For an elastic tracker that is based on two collisions, the intrinsic efficiency $\epsilon_2 \sim (\sigma_{el} n L)^2$. MWPCs and LFI can be used as absorptive trackers with $^3$He, $^6$Li and $^{10}$B. Some other atoms have to added to a gas mixture so that photons with appropriate wavelengths and sufficient intensities are available for LFI and time measurements. Atoms or compounds of $^3$He, $^6$Li, $^{10}$B, together with $\alpha$ sources such as $^{241}$Am, will be useful to developing LFI technique.

Detector pressures here are in the range of 1 to 10 bar STP. Although higher pressures or even solid density trackers may be desirable for higher intrinsic efficiency, higher densities may not be compatible with the MWPC operation. In LFI, solid density would require optics with sub-$\mu$m spatial resolution, which may compound the technology development in the beginning. Another difficulty is the apparent lack of lense material to focus VUV photons such as He I 58.4 nm. Initial experiments with $^1$H or $^2$H is easier, when both the visible line H I  656.3 nm and VUV line H I 121.6 nm are available. Other wavelengths may also be possible by using different gas mixtures, although we have only discussed H$_2$, CH$_4$ and their deuterated versions here.

Spatial resolution is limited by ion range straggling. For individual ions, however, the ultimate spatial resolution of ion tracking is limited to the primary ion track width, or the ranges of multiply scattered electrons as given by $R_e = 9.93 \times 10^{-6} E_e$ [g$\cdot$cm$^{-2}$keV$^{-1}$], independent of stopping medium. For electron energies $E_e $ = 1 to 4 keV, $R_e$ = 1 to 4 mm in 1 bar STP hydrogen gas (mass density 8.93 $\times$ 10$^{-5}$ g/cm$^{-3}$). At 10 bar pressure, the range shrinks by a factor of 10. In other gas mixtures, such as CH$_4$, the primary ion track width will be smaller by a factor of 7 from that of hydrogen. In all cases, the ion track widths are comparable or less than the range straggling of ions with energy above 1 MeV.

One possible application of the fast neutron trackers is to pin-point the locations of unknown neutron sources. Compared with a recoil proton telescope~\cite{Knoll2000,Mascarenhas:2009}, the proposed tracks have better angular resolution. Furthermore, when the TOF of neutrons can be measured and therefore two recoil events are sufficient for each fast neutron, the intrinsic efficiencies of the proposed trackers will be comparable to a recoil proton telescope with a similar areal density. A significant reduction of neutron events may be possible in order to pin-point an unknown neutron source using the proposed trackers.

\section{Conclusions}
We have discussed the principle of fast neutron tracking based on elastic collisions. The linear momentum of a fast neutron can be measured from as few as two and a `half' consecutive recoil ion tracks. If the time-delay between two consecutive recoil ion tracks is also measured, the number of ion track measurements can be reduced to one and a half. The magnitude and angular resolution of neutron momentum are limited by ion range straggling and ultimately by ion track width in the stopping medium. The angular resolution and magnitude of neutron moment measurement are about ten percent due to ion range straggling. Multi-wire proportional chambers and light-field imaging are proposed for fast neutron tracking. Single-charge or single-photon detection sensitivity would be required. Light-field imaging is free of charge-diffusion-induced image blur, but the limited number of photons available can be a challenge. $^1$H, $^2$H and $^3$He would be used for the initial demonstration of fast neutron trackers based on light-field imaging. 

\begin{acknowledgments}
We are thankful to Dr. David K. Wehe of the University of Michigan, Ann Arbor, Michigan for suggestions. 
\end{acknowledgments}


\bibliographystyle{elsarticle-num}
\bibliography{NeutronTrack3Ub}







\end{document}